\documentclass[12pt]{iopart}

\usepackage{iopams}
\usepackage{color}
\usepackage[dvips]{graphicx}  
\usepackage{epsfig}           

\newcommand{\HKane}{H_{\mathrm{K}}}
\newcommand{\Hkzero}{H_0}

\newcommand{\Hpert}{H'}
\newcommand{\Heff}{\tilde{H}}

\renewcommand{\vec}[1]{\mathbf{#1}}
\newcommand{\op}[1]{\hat{#1}}

\newcommand{\braket}[2]{\langle#1|#2\rangle}

\newcommand{\trevop}{\mathcal{T}}
\newcommand{\efield}{\mathcal{E}}
\newcommand{\vecsigma}{\boldsymbol \sigma}

\newcommand{\fAI}{f_{\mathrm{E+,1}}}
\newcommand{\fAIV}{f_{\mathrm{E+,4}}}
\newcommand{\fBIII}{f_{\mathrm{H+,3}}}
\newcommand{\fDVI}{f_{\mathrm{H-,6}}}
\newcommand{\fCII}{f_{\mathrm{E-,2}}}
\newcommand{\fCV}{f_{\mathrm{E-,5}}}
\newcommand{\fLHpI}{f_{\mathrm{LH+,1}}}
\newcommand{\fLHmII}{f_{\mathrm{LH-},2}}
\newcommand{\fLHpIV}{f_{\mathrm{LH+,4}}}
\newcommand{\fLHmV}{f_{\mathrm{LH-,5}}}

\newcommand{\fHHIIpIII}{f_{\mathrm{HH2+,3}}}
\newcommand{\fHHIImVI}{f_{\mathrm{HH2-,6}}}

\newcommand {\bra} [1] {\langle #1 |}
\newcommand {\ket} [1] {| #1 \rangle}

\begin{document}
\title[Fingerprint of Different Spin-Orbit Terms for Spin Transport in HgTe QWs]{Fingerprint of Different Spin-Orbit Terms for Spin Transport in HgTe Quantum Wells}
\author{D. G. Rothe$^1$, R. W. Reinthaler$^1$, C.-X. Liu$^{1,2}$, L. W. Molenkamp$^2$, S.-C. Zhang$^3$ and E. M. Hankiewicz$^1$}
\address{$^1$Institut f\"ur Theoretische Physik und Astrophysik, Universit\"at W\"urzburg, 97074 W\"urzburg, Germany}
\address{$^2$Physikalisches Institut (EP3), Universit\"at W\"urzburg, 97074 W\"urzburg, Germany}
\address{$^3$Department of Physics, McCullough Building, Stanford University, Stanford, CA 94305-4045}

\ead{hankiewicz@physik.uni-wuerzburg.de}

\begin{abstract}
Using $\vec{k}$$\cdot$$\vec{p}$ theory, we derive an effective four band model
describing the physics of  the
typical two-dimensional topological insulator (HgTe/CdTe quantum well) in the presence of
out-of-plane in z-direction inversion breaking and in-plane confining potentials.
We find that up to third order in perturbation theory,
only the inversion breaking potential generates new elements to the four band Hamiltonian that are off-diagonal in spin space.
When this new effective Hamiltonian is folded into an effective two band model for the conduction (electron)
or valence (heavy hole) bands, two competing terms appear:
(1) a Rashba spin-orbit interaction originating from
 inversion breaking potential in z-direction and
(2) an in-plane Pauli term as a consequence of the in-plane confining potential.
Spin transport in the conduction band  is further analysed within the Landauer-B\"uttiker formalism.
We find that for asymmetrically doped HgTe quantum wells, the behaviour of the spin-Hall conductance is dominated by the Rashba term.

\end{abstract}

\maketitle

\section{Introduction}
Two-dimensional (2D) topological insulators form a new state of  matter where the gapped bulk state is accompanied by gapless spin edge states
that are protected against small perturbations by time reversal symmetry \cite{Bernevig06,Koenig07,Koenig08,Roth09}.
In contrast to the chiral nature of the Quantum Hall state, where only one channel propagates at the edge of the system,
these new edge states have a helical nature, i.e. there are two counter-propagating spin-edge channels at the edge \cite{Bernevig06,Koenig07,Koenig08,Roth09}.
The topological protection of the edge states is connected with Kramer's theorem and states that a system with an
odd number of Kramer's pairs at a given edge is protected against single-particle excitations \cite{Koenig08}.
The simplest way to find topologically nontrivial insulators is to look at systems where the conduction and the valence bands
have  opposite parity and a change in band ordering (band inversion) occurs as a function of a tuning parameter like the strength of spin-orbit coupling \cite{Bernevig06,Fu07}.
This criterion leads to the unified Dirac form of the effective Hamiltonian for topological insulators with a spatial inversion center \cite{Bernevig06,Zhang09}.
In this paper we will be particularly interested in HgTe/CdTe quantum wells (QWs), which is the first topological insulator discovered
in nature\cite{Bernevig06,Koenig07}. This system can be tuned from the normal to 2D topological insulator phase by changing the thickness $d$ of the HgTe layer \cite{Bernevig06,Koenig07}.
Recent conductance measurements in multi-terminal structures \cite{Koenig07,Roth09} clearly show the existence of one-dimensional helical
edge channels in this material for $d$ larger than the critical value $d_c=6.3\,\mathrm{nm}$ \cite{Koenig07,Roth09}.

An effective four band model introduced by Bernevig, Hughes and Zhang (BHZ)\cite{Bernevig06}
consists of two disconnected blocks, each having the form of the Dirac Hamiltonian in 2D
and additional quadratic terms crucial for defining the concept of band inversion.
The BHZ model adequately describes the insulating regime in HgTe/CdTe QWs close to the $\Gamma$ point
and the topological quantum phase transition near the critical thickness $d=d_c$. It has been
extended to include the bulk inversion symmetry breaking effects in Ref. \cite{Koenig08}.
However, this model does not yet include the structural inversion asymmetry (SIA)
terms that can be very large in this narrow gap material. Indeed, it was shown experimentally that
an external top gate applied to the HgTe/CdTe QWs can change the energy of the Rashba spin-orbit splitting
in the range from 0 to 30meV \cite{Novik05} and the samples can be tuned from insulating to metallic regime \cite{Bruene08}.
Furthermore, the Aharonov-Casher oscillations \cite{Koenig06} as well as the ballistic spin-Hall effect in HgTe/CdTe QWs \cite{Bruene08},
which occur in the metallic regime, can be well described by an
effective two band (electron or heavy hole) model taking into account the Rashba
spin-orbit interactions. Therefore, it is desirable to build an unified Hamiltonian
for 2D topological insulators that includes
information about both the band structure and the presence of
an inversion breaking potential. This is the purpose of this paper.
Using  $\vec{k}$$\cdot$$\vec{p}$ theory we derive an extension of the BHZ model
describing a 2D topological insulator which includes an  out-of-plane ( in z direction) inversion breaking potential and an in-plane confining potential.
The central result of this paper, the generalized  four band Hamiltonian, is presented in equation \eref{hgte2deg}
and should also be applicable to other 2D topological insulators, such as type II InAs/GaSb quantum wells\cite{Liu08} and Bi$_2$Se$_3$ thin films \cite{Liu10,Linder10,Lu2009}.
Next, we will use the Foldy-Wouthuysen transformation to find an effective model describing
electron or heavy hole band.
We show that such an effective
model contains two different types of spin-orbit interactions,
one of them is the well-known Rashba spin-orbit interaction induced by
the inversion breaking potential in z-direction, while the other originates from the in-plane confining potential,
and is referred to as the in-plane Pauli term.
Although both of these terms, for the conduction band, are linear in the wave vector and the spin,
they contribute differently to the spin transport.
The first (Rashba) term does not conserve the z-component of spin, $S_z$, causing spin precession, while the in-plane Pauli term conserves $S_z$.
We study the interplay of the Rashba and in-plane Pauli terms.
We predict that the spin-Hall conductance will show the precession pattern as a function
of the inversion breaking potential in z-direction even in the presence of a strong confining potential.
Further, the  strong in-plane confining potential enhances  the spin-Hall conductance generated by the Rashba term, because it partially fixates the direction of the precessing spin.
Therefore the behaviour of the spin transport in asymmetrically doped quantum wells should be dominated by the Rashba term and it is justified to describe the spin-Hall conductance in the metallic regime through simple
effective models for electrons and heavy holes
(see equations \eref{hcondband} and \eref{hvalband} of this manuscript), as long as the band gap is non-zero.

The rest of the article is organized as follows. Section 2.1 gives the derivation of the
effective four band model for a 2D topological insulator with a spatial inversion breaking potential using $\vec{k}$$\cdot$$\vec{p}$ theory.
In Section 2.2 we show that the same Hamiltonian can be derived using general symmetry arguments.
In Section 3 we derive an effective one band model with competing Rashba-type and in-plane Pauli
contributions. Section 4 describes the interplay between both terms within the
Landauer-B\"uttiker formalism. We finish the paper with conclusions.

\section{Effective Hamiltonian for HgTe QWs in the presence of the inversion
breaking potential in z-direction and the in-plane confining potential.}

\subsection{Derivation of the extended HgTe Hamiltonian within $\vec{k}$$\cdot$$\vec{p}$ theory.}

In this section we will consider the influence of the
structural inversion asymmetry on HgTe/CdTe quantum wells
(QWs) and derive a corresponding effective $4\times4$ model
 with an out-of plane (in z-direction) inversion breaking potential.
Our starting point is the eight-band Kane Hamiltonian $\HKane$, which is
described in \cite{Novik05}. The spin-orbit split-off bands
$\ket{\Gamma_7,\pm 1/2}$ are far away in energy from the other bands
and are not important for the description of the quantum well.
This is in contrast to the bulk case, where SIA terms are nonzero only when the $\Gamma_7$
band is taken into account \cite{Winkler05}.

Therefore we limit ourself to the upper $6\times6$ block
of the Kane model with the basis set of wave functions in the sequence
$(\ket{1}=\ket{\Gamma^-_6, 1/2}, \ket{2}=\ket{\Gamma^-_6, -1/2},$
$\ket{3}=\ket{\Gamma^+_8, 3/2}, \ket{4}=\ket{\Gamma^+_8, 1/2},$
$\ket{5}=\ket{\Gamma^+_8, -1/2},$ $\ket{6}=\ket{\Gamma^+_8, -3/2})$
where we use the standard notation with $\ket{\Gamma^-_6, \pm 1/2}$ describing s-like conduction band,  $\ket{\Gamma^+_8, \pm 1/2}$ p-like light hole band
 and  $\ket{\Gamma^+_8, \pm 3/2}$ p-like heavy hole band in zincblende crystal structures\cite{Winkler05}.

In the following we always use $\ket{\alpha}$ ($\alpha=1,2,\dots, 6$)
to denote the basis set of wave functions shortly.
We consider a quantum well configuration with HgTe layers
sandwiched by two CdTe barrier layers along z direction,
hence the parameters of the Kane model $\HKane$ have spatial dependence \cite{Bernevig06}.
The matching of wave functions in z-direction for HgTe/CdTe QWs has to be done very carefully
because the bulk barrier material CdTe has a normal band structure with  $\ket{\Gamma^-_6, \pm 1/2}$ above $\ket{\Gamma^+_8}$ bands
while bulk HgTe has inverted band ordering with $\ket{\Gamma^+_8}$ above $\ket{\Gamma^-_6, \pm 1/2}$  bands \cite{Bernevig06,Koenig07}.
This is exactly the reason for the change from the normal to inverted band structure ordering for the HgTe/CdTe QWs above the critical value of HgTe layer width $d_c= 6.3$nm (see for example Fig.~1 in \cite{Koenig07}).

The envelope function approximation \cite{Burt88} is applied to
solve the eigenproblem of the quantum well. Since the Kane model preserves
inversion symmetry, in order to discuss the SIA, we need to
take into account an additional potential $V({\mathbf r})
= V_0(x,y) + z e\efield_z$,
where $e>0$ is the elementary charge and $z e\efield_z$ is
 the inversion breaking potential in z-direction,
while $V_0(x,y)$ is the in-plane confining potential and its form will be specified in section 4. Then the full Hamiltonian is
\begin{eqnarray}
	\hat{H}_{\mathrm{full}}=H_{\mathrm{K}}(k_\parallel,z) + V({\mathbf r}).
	\label{eq:Hamfull}
\end{eqnarray}
Next we split the Hamiltonian (\ref{eq:Hamfull}) into two parts
$\hat{H}_{\mathrm{full}} =  \Hkzero + H'$, where $\Hkzero$ is the Kane Hamiltonian
when $k_\|=0$ and is treated as the zero-order Hamiltonian. Explicitly,
$\Hkzero$ is given by
\newcommand{\Pkz}{\sqrt{\frac{2}{3}} P \op{k}_z}
\begin{eqnarray}
\label{kanekzero}
\fl \Hkzero = H_{\mathrm{K}}(k_{\|} = 0) =
\left(
\begin{array}{cccccc}
  T^{(0)} 	& 	0 	& 	0 	&  \Pkz & 0 & 0 \\
 0 		& T^{(0)}	&	0 	& 0 & \Pkz & 0 \\
 0 		& 0 		&  W^{(0)}_+  & 0 & 0 & 0 \\
 \Pkz & 0 & 0 & W^{(0)}_- & 0 & 0 \\
 0 & \Pkz & 0 & 0 & W^{(0)}_- & 0 \\
 0 & 0 & 0 & 0 & 0 & W^{(0)}_+
\end{array}
\right)
\end{eqnarray}
where $\hat{k}_z$ is an operator and the heavy hole bands ($\Gamma_8^{+},\pm 3/2$) are completely decoupled
from the electron and light hole bands. Here
$P = -\frac{\hbar^2}{2 m_0} \braket{S}{p_x | X}$
is the Kane matrix element between the $\Gamma_6$ and $\Gamma_8$ bands,
while the other parameters are given by
\begin{eqnarray}
T^{(0)} &=&  E_c(z)  +  \frac{\hbar^2 }{2 m_0}\op{k}_z (2 F(z) + 1) \op{k}_z \\
W^{(0)}_{\pm} &=& E_v(z) + \frac{\hbar^2}{2 m_0} \op{k}_z \left(2 \gamma_2(z) \mp \gamma_1(z) \right) \op{k}_z
\end{eqnarray}
with $F(z)  =  \frac{1}{m_0} \sum_{j}^{\Gamma_5}
\frac{\left|\braket{S}{p_x|u_j}\right|^2}{E_c(z) - E_j(z)}$
including remote bands $\ket{u_j}$ with $\Gamma_5$ symmetry perturbatively.
$E_{c/v}$ designate the positions of  the conduction/valence band edges and the $\gamma_i$ are renormalized
Luttinger parameters \cite{Luttinger56}. The axial approximation is adopted
\cite{Ekenberg85, Jeschke00} in order to keep the in-plane
rotation symmetry.

$H'$ is treated as a perturbation, and is written as
\begin{eqnarray}
\label{kaneprime}
\fl H' = \HKane(k) - \Hkzero + V =
\left(
\begin{array}{cccccc}
 T^{(1)}  	&	 0 & -\frac{P k_+}{\sqrt{2}} & 	0 	& \frac{P k_-}{\sqrt{6}} 	& 	0 \\
 0 	& T^{(1)} & 0 	& -\frac{P k_+}{\sqrt{6}} 	& 0 	& \frac{P k_-}{\sqrt{2}} \\
 -\frac{P k_-}{\sqrt{2}} & 0 	& W^{(1)}_+  & -S_- & R & 0 \\
 0 & -\frac{P k_-}{\sqrt{6}} & -S_-^{\dagger}  & W^{(1)}_-  & C  & R \\
 \frac{P k_+}{\sqrt{6}} & 0 & R^{\dagger} & C^{\dagger} & W^{(1)}_-  & S_+^{\dagger} \\
 0 & \frac{P k_+}{\sqrt{2}} & 0 & R^{\dagger} & S_+ & W^{(1)}_+
\end{array}
\right)
\end{eqnarray}
with $k_{\pm} = k_x \pm i k_y$,
$C = \frac{\hbar^2}{2 m_0} k_- [\kappa,k_z]$,
$R = \frac{\sqrt{3}\hbar^2}{2 m_0} \bar{\gamma} k_-^2$,
$S_{\pm} = -\frac{\sqrt{3} \hbar^2}{2 m_0} k_{\pm} \left(\left\{\gamma_3,k_z\right\} + [\kappa,k_z]\right)$,
$T^{(1)} =  \frac{\hbar^2 (2 F+1) k_{\|}^2}{2 m_0} + V$ and
$W^{(1)}_{\pm} = - \frac{\hbar^2}{m_0} \left(\gamma_1 \pm  \gamma_2\right) k_{\|}^2 + V$.
Here, $\bar{\gamma} = (\gamma_3 + \gamma_2)/2$.
$\kappa$ is the renormalized Luttinger parameter related to the  part of Hamiltonian which is antisymmetric in the components of $\vec{k}$.
In the original Luttinger model, it was introduced because in the presence of a
magnetic field the components of $\vec{k}$ do not commute. In our case, it appears because the material parameters are functions of the $z$ coordinate.

Now we  will  generalize the BHZ approach \cite{Bernevig06}
to project the Hamiltonian (\ref{eq:Hamfull}) into the low energy
sub-space, which can be done in two steps. First, we numerically
diagonalize the Hamiltonian $H_0$, so that
$H_0|i\rangle=E_i|i\rangle$, to obtain the eigenenergies $E_i$ and eigenstates $|i\rangle$ of the quantum well.
Here the eigenstate $|i\rangle$ can be expanded
in the basis $|\alpha\rangle$ as $\ket{i}=\sum_{\alpha} f_{i,\alpha}(z)\ket{\alpha}$, where
the function $f_{i}(z)$ gives the envelope function along z-direction
for the quantum well. We use Greek indices to indicate basis functions of the Kane Hamiltonian and Roman
indices to denote the subbands.
The envelope function components $f_{i,\alpha}(z)$
are calculated with the help of the numerical diagonalization of $H_0$.

In order to perform the degenerate perturbation calculation, we need
to cast the eigenstates of $H_0$ into two classes. The first one, denoted as class A,
includes the basis wave functions of our final four band effective model.
As shown by BHZ \cite{Bernevig06}, for HgTe/CdTe quantum wells,
it is necessary to take into account the two electron-like subbands
$\ket{E1,\pm}$ and two heavy hole subbands $\ket{H1,\pm}$,
which are expanded explicitly as
\begin{eqnarray}
\label{newbasis}
|E1,+\rangle & = &  \fAI(z) |1\rangle + \fAIV(z) |4\rangle  \\
|H1,+\rangle  & = & \fBIII(z) |3\rangle \\
|E1,-\rangle  & = & \fCII(z) |2\rangle + \fCV(z) |5\rangle \\
|H1,-\rangle  & = & \fDVI(z) |6\rangle
\end{eqnarray}
As pointed out above, for $H_0$ the heavy hole bands are decoupled from
the electron and light hole bands, therefore the eigenstate
$\ket{H1,+(-)}$ consists only of the basis $\ket{3}$ ($\ket{6}$)
while $\ket{E1,+(-)}$ is a combination of the basis $\ket{1}$ ($\ket{2}$)
and $\ket{4}$ ($\ket{5}$). The second class, denoted as class B, includes the states
which need to be taken into account in the following perturbation procedure.
Here we consider the first light hole-like subbands $\ket{LH,\pm}$
and the second and third heavy hole subbands $\ket{HH2,\pm}$ and $\ket{HH3,\pm}$, which are
written explicitly as
\begin{eqnarray}
|LH,+\rangle   & = & \fLHpI(z) |1\rangle +  \fLHpIV(z) |4\rangle  \\
|HHn,+\rangle  & = & f_{\mathrm{HHn+,3}}(z) |3\rangle \\
|LH,-\rangle  & = &  \fLHmII(z) |2\rangle + \fLHmV(z) |5\rangle \\
|HHn,-\rangle  & = & f_{\mathrm{HHn-,6}}(z) |6\rangle.
\end{eqnarray}
All the other subbands of the quantum well are neglected here
since they are well separated in energy.

Before we go to the next step of the perturbation calculation,
it is useful to have a look at the symmetry properties
of the relevant states. For Hamiltonian $H_0$,
we have three types of symmetries: the time reversal
symmetry $\trevop$, the inversion symmetry $P$ and the
in-plane full rotation symmetry $R_z(\theta)$.
For the time
reversal operation $\trevop$, it is not hard to show
that $|E1,\pm\rangle$ ($|H1,\pm\rangle$) are Kramer's partners, i.e.
$\trevop |E1,+\rangle =  |E1,-\rangle$,
$\trevop |E1,-\rangle = -|E1,+\rangle$,
$\trevop |H1,+\rangle =  |H1,-\rangle$ and
$\trevop |H1,-\rangle = -|H1,+\rangle$, and we use phase conventions
for the envelope functions which yield the relations
$\fCII = \fAI^*$, $\fCV = -\fAIV^*$, $\fDVI = \fBIII^*$, $\fLHmII = -\fLHpI^*$
and $\fHHIImVI = -\fHHIIpIII^*$.
Time reversal symmetry relates states
with opposite spin to each other, hence when the effective
Hamiltonian for one spin is constructed,
the Hamiltonian for the opposite spin can be
easily obtained through the operation $\trevop$.
The inversion operation $P$ defines the parity of each
subband, which can greatly simplify the matrix elements
in the perturbation procedure below. The parity of the
subbands $\ket{i}$ in the quantum well is determined
by both the envelope function $f_{i,\alpha}(z)$ and the basis
wave function $\ket{\alpha}$. The parities of the envelope functions
can be obtained through numerical calculation \cite{Bernevig06, Jeschke00},
and are listed in \tref{tabfenvparities}.
\begin{table}[h]
 \caption{\label{tabfenvparities}Parities of the envelope function components.}	
 \begin{indented} \item[]
  \begin{tabular}{c|cccccccc}
  \hline
   even: & $\fAI$ & $\fCII$ & $\fLHpIV$ & $\fLHmV$  & $\fBIII$ & $\fDVI$ & $f_{\mathrm{HH3+,3}}$ & $f_{\mathrm{HH3-,6}}$  \\
\hline
   odd:  & $\fAIV$ & $\fCV$ & $\fLHpI$ & $\fLHmII$ &  $\fHHIIpIII$ & $\fHHIImVI$ & &\\
    \hline
 \end{tabular}
 \end{indented}
\end{table}
The parities of the basis functions are given by $P|\Gamma^-_6,\pm 1/2\rangle=-|\Gamma^-_6,\pm 1/2\rangle$,
$P|\Gamma^+_8,\pm\rangle=+|\Gamma^+_8,\pm\rangle$.
 Thus the parities of the subbands are $P\ket{E1\pm}=-\ket{E1\pm}$, $P\ket{H1\pm} = \ket{H1\pm}$, $P\ket{LH\pm} = \ket{LH\pm}$, $P\ket{HH2\pm} = -\ket{HH2\pm}$ and $P\ket{HH3\pm} = \ket{HH3\pm}$.
Due to the in-plane rotation symmetry, the total angular momentum $J$ along z-direction
is a good quantum number, which can be used to identify the eigenstates. Since the electron-like subbands
have  $J=\frac{1}{2}$, the rotation operator is $R_z(\theta)\ket{E1\pm}=e^{\pm i\frac{\theta}{2}}\ket{E1\pm}$
while for the heavy hole subbands with $J=\frac{3}{2}$, it should be $R_z(\theta)\ket{H1\pm}
=e^{\pm i\frac{3\theta}{2}}\ket{H1\pm}$.

Next, we calculate the effective Hamiltonian of the four states
in the class A based on quasi-degenerate perturbation theory. All states in classes A and B
are eigenstates of Hamiltonian $H_0$. However when $H'$ is introduced, they are no longer
eigenstates due to mixing between the states of class A and class B. Therefore,
treating $H'$ as a small perturbation, we need to perform an unitary transformation
to eliminate the coupling between the states in class A and class B up to the required order.
Details of the perturbation procedure can be found in \cite{Winkler05},
and here we directly apply the following third order perturbation formula:
\newcommand{\hspx}{\hspace{0cm}}
\begin{eqnarray}
\label{perturbationformula}
\fl \hspx \tilde{H}_{mm'}  =  E_m\delta_{mm'} + \Hpert_{mm'}
+ \frac{1}{2} \sum_l \Hpert_{ml} \Hpert_{lm'} \left(\frac{1}{E_m - E_l} + \frac{1}{E_{m'} - E_l}\right)
\nonumber\\
\fl \hspx  - \frac{1}{2} \sum_{l,m''} \left( \frac{ \Hpert_{ml} \Hpert_{lm''} \Hpert_{m''m'}}{(E_{m'} - E_l ) (E_{m''} - E_l)} + \frac{\Hpert_{mm''} \Hpert_{m''l} \Hpert_{lm'}} {(E_m - E_l) (E_{m''} - E_l)}\right)
\nonumber \\
\fl \hspx + \frac{1}{2} \sum_{l,l'} \Hpert_{ml} \Hpert_{ll'}\Hpert_{l'm'} \Big(\frac{1}{(E_m - E_l) (E_m - E_{l'})} + \frac{1}{(E_{m'} - E_l ) (E_{m'} - E_{l'})} \Big)
\end{eqnarray}
with
\begin{eqnarray}
	{H'}_{jk} := \braket{f_j}{H' | f_k} =
\int dz \, \sum_{\alpha,\beta = 1}^{6} f_{j,\alpha}^{*}(z) (H')_{\alpha \beta} \,  f_{k,\beta}(z).
	\label{scalarprod}
\end{eqnarray}
The summation indices $m,m',m''$ are taken from the states in class $A$ ($E1+,E1-, H1+, H1-$)
and indices $l,l'$ are from the states in class $B$ ($LH+,LH-, HH2+, HH2-, HH3+, HH3-$). As mentioned above, Greek indices denote
entries of the Kane matrix \eref{kaneprime}.
Here we should keep in mind that the order of the matrix elements
of $H'$ in (\ref{perturbationformula}) is important,
as they may not commute with each other.

The perturbation calculation based on (\ref{perturbationformula}) is straight-forward but lengthy.
The parities of the envelope functions discussed above can be used to reduce the number of the
matrix elements of $H'$. For example, the first-order term
\begin{eqnarray}
\fl \bra{E1+} H' \ket{E1-} =  -\frac{P k_+}{\sqrt{6}}\braket{\fCII}{\fAIV} + \frac{P k_+}{\sqrt{6}}\braket{\fCV}{\fAI} + \bra{\fCV}C^{\dagger}\ket{\fAIV}
\end{eqnarray}
vanishes completely because all integrands are odd functions of $z$ (see \tref{tabfenvparities}).

In the four-band basis $(\ket{E1+}, \ket{H1+}, \ket{E1-},  \ket{H1-})$,
the final effective Hamiltonian is written as
\begin{eqnarray}
\label{hgte2deg}
\Heff = \tilde{H}_0 + \tilde{H}_{\mathrm{R}} + V_0(x,y) \\
\label{H0}
\tilde{H}_0=
 \epsilon(k)I + \left(\begin{array}{cccc}
  \mathcal{M}(k)  	&  \mathcal{A} k_+  	 	& 0	  & 0   \\
   \mathcal{A} k_-   		&  -\mathcal{M}(k)  	& 0	  & 0  \\
   0	 	& 0	       	& \mathcal{M}(k)  & - \mathcal{A} k_-	\\
  0	   	& 0	 	&  -\mathcal{A} k_+ 	  & -\mathcal{M}(k)
  \end{array}\right)\\
 \label{HR}
 \tilde{H}_{\mathrm{R}}=
 \left(\begin{array}{cccc}
  0  	&  0	& -i R_0 k_-	  & -S_0 k_-^2   \\
   0	&  0	& S_0 k_-^2	  & i T_0 k_-^3  \\
   i R_0 k_+	 	&  S_0 k_+^2	       	& 0  & 0	\\
  -S_0 k_+^2	   	& -i T_0 k_+^3	 	&  0	  & 0
  \end{array}\right)
\end{eqnarray}
with $\mathcal{A} = A + A_2 k^2$, $\mathcal{M}(k)= M -B k^2$,$\epsilon(k) = C - D k^2$ and $I$ is diagonal unit matrix. Here
$k^2 = k_x^2 + k_y^2$.

We note that $\tilde{H}_0$ is equivalent to the BHZ Hamiltonian in
\cite{Bernevig06} if we further omit the k-dependence of the $\mathcal{A}$, by setting $A_2=0$. We also assume that the reference energy is fixed in the middle of the gap, i.e. $C=0$.
Besides the BHZ Hamiltonian, we find a new term $\tilde{H}_{\mathrm{R}}$, which is off-diagonal
in spin space due to the inversion breaking potential $ze\efield_z$.
As mentioned above we have included the subbands $|HH3,\pm\rangle$ in the calculation of the effective parameters A,B,D. However these subbands do not contribute to $\tilde{H}_{\mathrm{R}}$. This is a consequence of the fact that the envelope functions belonging to $|HH3,\pm\rangle$ have parities opposite to the envelope functions of $|HH2,\pm\rangle$, see \tref{tabfenvparities}.

There are three new terms in $\tilde{H}_{\mathrm{R}}$. The first term ($R_0$ term)
originates from the second order perturbation theory and is exactly the electron Rashba term with
\begin{eqnarray}
	\label{eq:parr_R0}
R_0  =
\frac{ie\efield_z}{3 (E_{\mathrm{LH}}-E_{\mathrm{E1}})}  \Bigg[
 \left( \bra{\fAI} z \ket{\fLHpI}
     + \bra{\fAIV} z \ket{\fLHpIV}
   \right)\cdot
\\
\fl \left( \sqrt{6} P  \braket{ \fLHpIV } { \fAI^* }
       + \sqrt{6} P \braket{ \fLHpI } { \fAIV^* }
       + \frac{3 \hbar^2} {m_0} \bra{\fLHpIV}[\kappa ,k_z] \ket{\fAIV^*}
 \right)
\nonumber \Bigg].
\end{eqnarray}
Here $\langle f_{i,\alpha}| O |f_{j,\beta} \rangle=\int dz
f^*_{i,\alpha}(z) O f_{j,\beta}(z)$ for an arbitrary operator $O$.
The electron Rashba term is linear in $k$ because of the $\frac{1}{2}$
electron spin.

The second term ($T_0$ term) originates from the third order perturbation
and denotes the heavy-hole $k^3$ Rashba term with the parameter
\begin{equation}
 T_0 =
\frac{-i\sqrt{3} e\efield_z \hbar^2}{8 m_0^2} \cdot (s_1 + s_2 + s_3)
\end{equation}
with
\begin{footnotesize}
\begin{eqnarray}
 \label{eq:selements}
\fl s_1 =
\frac{2}{ (E_{\mathrm{E1}} - E_{\mathrm{HH2}}) ( E_{\mathrm{H1}} - E_{\mathrm{HH2}})}
\Bigg[ \left(\sqrt{2} P m_0 \braket{\fBIII}{\fAI}
             -\sqrt{3}\hbar^2 \bra{\fBIII} (\left\{\gamma _3,k_z\right\} + [\kappa ,k_z]) \ket{\fAIV}
        \right)
  \nonumber \\
   \cdot \bra{\fAIV} \bar{\gamma} \ket{\fHHIIpIII^*} \bra{\fHHIIpIII^*} z \ket{ \fBIII^*} \Bigg]
\\
\fl s_2 =
\frac{2} {(E_{\mathrm{E1}} - E_{\mathrm{LH}})(E_{\mathrm{H1}} - E_{\mathrm{LH}})} \Bigg[
   \left( \sqrt{2} P m_0 \braket{\fBIII}{\fAI}
               -\sqrt{3}\hbar^2 \bra{\fBIII}
		(\left\{\gamma_3, k_z\right\} + [\kappa, k_z ]) \ket{\fAIV}
	\right) \nonumber \\
        \cdot \left(  \bra{\fAI} z \ket{\fLHpI}
                + \bra{\fAIV} z \ket{\fLHpIV}  \right)
	\bra{\fLHpIV}\bar{\gamma} \ket{\fBIII^*} \Bigg]
\\
\fl s_3 =
 \frac{4}{(E_{\mathrm{H1}} - E_{\mathrm{HH2}}) (E_{\mathrm{H1}} - E_{\mathrm{LH}})} \Bigg[
  \bra{\fBIII} z \ket{\fHHIIpIII}
  \nonumber \\
\fl  \cdot
  \left( \sqrt{2} P m_0 \braket{ \fHHIIpIII}{ \fLHpI}
	       - \sqrt{3} \hbar^2 \bra{\fHHIIpIII}
		(\left\{\gamma_3,k_z \right\} + [\kappa,k_z]) \ket{\fLHpIV}
	  \right)
  \bra{\fLHpIV} \bar{\gamma} \ket{\fBIII^*} \Bigg]
\end{eqnarray}
\end{footnotesize}
Since for heavy holes the spin is $3/2$, the change of angular momentum
upon a spin-flip is $3$, which corresponds to $k_{\pm}^3$.

The third term ($S_0$ term), which is proportional to $k^2$, also comes
from the second order perturbation with the parameter
\begin{eqnarray}
\fl S_0 =
-\frac{\sqrt{3}\hbar^2  e\efield_z }{4 m_0} \Bigg[
 \left(\frac{1}{E_{\mathrm{E1}}-E_{\mathrm{HH2}}} + \frac{1}{E_{\mathrm{H1}}-E_{\mathrm{HH2}}} \right)
 \bra{\fBIII} z \ket{ \fHHIIpIII }
 \bra{\fHHIIpIII} \bar{\gamma } \ket{\fAIV^*}
\\
\fl
+
\left(\frac{1}{E_{\mathrm{E1}}-E_{\mathrm{LH}}}+\frac{1}{E_{\mathrm{H1}}-E_{\mathrm{LH}}}\right)
\bra{\fBIII} \bar{\gamma} \ket{\fLHpIV^*}
( \bra{\fLHpI^*} z \ket{\fAI^*}
  + \bra{\fLHpIV^*} z \ket{\fAIV^*})
  \Bigg] \nonumber.
\end{eqnarray}
This is a new off-diagonal term between the electron-like $\frac{1}{2}$
($-\frac{1}{2}$) and heavy-hole $-\frac{3}{2}$ ($\frac{3}{2}$) states
and the change of the angular momentum is 2, corresponding to $k_\pm^2$.
All parameters
can be determined by using the numerically calculated $f_{i,\alpha}(z)$, and are listed in the table \ref{tab:values}
for a QW with a thickness $d_0=\mathrm{70}$\AA.

We note that all three terms in the inversion breaking Hamiltonian $\tilde{H}_R$
are proportional to $\mathcal{E}_z$ but are independent of $V_0(x,y)$.
The corrections originating from $V_0(x,y)$ to the $2\times 2$ off-diagonal blocks in \eref{HR} are of higher order in the perturbation than the ones coming from $\efield_z$. Moreover, they must contain both $\efield_{\|} = \nabla V_0(x,y)/e$ and $\efield_z$ due to the fact that the
 in-plane field $\efield_{\|}$ does not break the $z\to-z$ inversion symmetry.
 Furthermore, $V_0(x,y)$ introduces corrections of third or higher order to the element $(\tilde{H}_{0})_{12}$ which have  the form
$[k_+,[k_+,V_0(x,y)]] k_-$ and $\nabla \vec{\efield_{\|}}\cdot k_+$. These corrections are much smaller than the element
$Ak_{+}$ which appears already in the first order of perturbation theory in $\tilde{H}_0$.
Corrections to the diagonal elements of $\tilde{H}_0$ induced by the in-plane potential are also very small.
The latter corrections, after folding to the electron or heavy hole subbands (see section 3),
produce similar contributions to the ones originating from $\tilde{H}_0$, but are an order of magnitude smaller due to a large energy separation between main bands and bands which are treated perturbatively.
Therefore, we justified that the only significant contribution to $\Heff$ connected with the in-plane potential comes from the bare diagonal potential $V_0(x,y)$ as shown in \eref{hgte2deg}.

\begin{table}
 \caption{\label{tab:values}Parameters of the effective $4 \times 4$ Hamiltonian, calculated for the quantum well width 70~\AA, at charge density $n = 2 \cdot 10^{10} \;\mathrm{cm}^{-2}$}
 \begin{indented}\item[]
\begin{tabular}{|l|r|}
\hline
$A \; [\mathrm{nm\; eV}]$   & $0.365$  \\
\hline
$B \; [\mathrm{nm^2\; eV}]$ & $-0.700$  \\
\hline
$D \; [\mathrm{nm^2 \;eV}]$ & $-0.525$  \\
\hline
$M \; [\mathrm{eV}]$ 	 & $-10.08 \cdot 10^{-3}$ \\
\hline
$R_0/(e\efield_z) \; [\mathrm{nm}^2]$  & $-15.6$ \\
\hline
$T_0/(e\efield_z) \; [\mathrm{nm}^4]$ & $-8.91$ \\
\hline
$i S_0 /(e\efield_z) \; [\mathrm{nm}^3] $  &  $-2.10$ \\
\hline
\end{tabular}
\end{indented}
\end{table}

\subsection{Symmetry arguments for the validity of the extended HgTe Hamiltonian.}
The goal of this subsection is to derive the effective 4$\times$4 Hamiltonian (\ref{hgte2deg}) using
the theory of invariants \cite{Winkler05}.
The theory of invariants states that the Hamiltonian must be invariant under all symmetry operations of
the considered system.
As discussed in the last section, the system has
time reversal symmetry $\trevop$, space inversion symmetry $P$ and
in-plane rotation symmetry $R_z(\theta)$. The transformation of
the set of basis wave functions under these symmetries for the effective model have been discussed
in the last section. The symmetry operations
in the matrix form for
the basis $(\ket{E1+}, \ket{H1+}, \ket{E1-},  \ket{H1-})$ are given by
\begin{itemize}
\item {\it Time reversal symmetry}: $\trevop = \Theta K$,
where $\Theta=-i\sigma_2 \otimes1$ and $K$ is the complex conjugate operator.
\item {\it Inversion symmetry}: $P= - 1\otimes\tau_3$
\item {\it Rotation symmetry}: $R_z(\theta)=e^{i\frac{\Sigma_z}{2}\theta}$
	with $\Sigma_z=\sigma_3\otimes(\frac{1+\tau_3}{2}+\frac{3(1-\tau_3)}{2})
	=\sigma_3\otimes(2-\tau_3)$
\end{itemize}
where the $\sigma_i$ denote Pauli matrices acting on the spin basis and the $\tau_i$ represent Pauli matrices acting
on the electron or heavy-hole subbands.

Generally, any four by four Hamiltonian can be expanded using
$\Gamma$ matrices as
\begin{eqnarray}
	\hat{H}_{\mathrm{eff}}=\epsilon({\bf k}){\bf I}+\sum_id_i({\bf k})\Gamma_i
	+\sum_{ij}d_{ij}({\bf k})\Gamma_{ij}
	\label{eq:sys_Heff}
\end{eqnarray}
where ${\bf I}$ is the $4\times4$ identity matrix, $\Gamma_i$ ($i=1\cdots 5$)
denote five $\Gamma$ matrices, which satisfy $\left\{ \Gamma_i,\Gamma_j
\right\}=2\delta_{ij}$, and the ten commutators of $\Gamma$ matrices
$\Gamma_{ij}=[\Gamma_i,\Gamma_j]/2i$.
$\epsilon({\bf k})$, $d_i({\bf k})$ and $d_{ij}({\bf k})$ can be expanded
as polynomials of the momentum ${\bf k}$. The Hamiltonian should be
invariant under the symmetry operations $P$, $\trevop$ and $R_z$, which indicates
that $d_{i(j)}(\mathbf{k})$ should behave the same as $\Gamma_{i(j)}$.
Therefore we need to work out the transformation form of the $d_{ij}({\bf k})$
and the $\Gamma$ matrices. We construct the $\Gamma$ matrices as follows
\begin{equation}
\begin{array}{lll}
\Gamma_1=\sigma_1\otimes\tau_1  &
    \Gamma_2=\sigma_2\otimes\tau_1 &
    \Gamma_3=\sigma_3\otimes\tau_1  \\
\Gamma_4=1\otimes\tau_2 &
    \Gamma_5=1\otimes\tau_3 &
    \Gamma_{ij}=\varepsilon_{ijk}\sigma_k\otimes1 \\
\Gamma_{i4}=\sigma_i\otimes\tau_3 &
    \Gamma_{i5}=-\sigma_i\otimes\tau_2 &
    \Gamma_{45}=1\otimes\tau_1
   \label{eq:sys_GamMat1}
¸\end{array}
\end{equation}
where $i,j=1,2,3$.
It is easy to prove that $(\Gamma_{ab})^2=1$, $\{\Gamma_a,\Gamma_{ab}\}=0$
and $\{\Gamma_{ab},\Gamma_{ac}\}=0$ for $b\neq c$.
For the above fifteen $\Gamma$ matrices, it is easy to calculate the
symmetry transformation under the time reversal operation $\trevop$ and
inversion operation $P$, which are listed in \tref{tab:symmetry}.
For the in-plane rotation operation $R_z(\theta)$, we can calculate
the transformation rule $\Gamma'(\theta)=e^{i\frac{\Sigma}{2}\theta}\Gamma
e^{-i\frac{\Sigma}{2}\theta}$ with the help of the differential equation
$\frac{d\Gamma'(\theta)}{d\theta}=\frac{i}{2}[\Sigma,\Gamma'(\theta)]$.
The obtained results are also given in table \ref{tab:symmetry}, from
which we find that $\Gamma_5$, $\Gamma_{12}$ and $\Gamma_{34}$
behave as a scalar under the rotation $R_z$,
$(\Gamma_4,\Gamma_3)$, $(\Gamma_{45},\Gamma_{35})$  and
$(\Gamma_{14}+\Gamma_{23},\Gamma_{31}+\Gamma_{24})$ rotate
as a vector with angular momentum 1, $(\Gamma_1,\Gamma_2)$,
$(\Gamma_{15},\Gamma_{25})$  corresponds to angular
momentum 2, and $(\Gamma_{23}-\Gamma_{14},\Gamma_{31}-\Gamma_{24})$
corresponds to angular momentum 3. In \tref{tab:symmetry}
we also list the corresponding tensors formed by $\mathbf{k}$
up to the order $k^3$.
\begin{table}[htb]
  \centering
  \begin{minipage}[t]{0.8\linewidth}
      \caption{ Summary of the symmetry properties of $\Gamma$ matrice and the tensors
      formed by $\mathbf{k}$. }
\label{tab:symmetry}
\begin{center}
\begin{tabular}{cccc}
    \hline\hline
      & $R_z$ & $\trevop$ & $P$ \\\hline\hline
      $\mathbf{I}$                 & 0 & + & + \\\hline
      $(\Gamma_1,\Gamma_2)$    & 2 & - & - \\\hline
      $(\Gamma_4,\Gamma_3)$    & 1 & - & - \\\hline
      $(\Gamma_{15},
      \Gamma_{25})$             & 2 & + & - \\\hline
      $(\Gamma_{45},
      \Gamma_{35})$             & 1 & + & - \\\hline
      $\Gamma_5$                 & 0 & + & + \\\hline
      $\Gamma_{12}$ &              0 & - & + \\\hline
      $\Gamma_{34}$ &              0 & - & + \\\hline
      $(\Gamma_{14}+\Gamma_{23},
      \Gamma_{31}+\Gamma_{24})$ & 1 & - & + \\\hline
      $(\Gamma_{23}-\Gamma_{14},
      \Gamma_{31}-\Gamma_{24})$ & 3 & - & + \\\hline\hline
      $(k_x,k_y)$              & 1 & - & - \\\hline
      $k_x^2+k_y^2$              & 0 & + & + \\\hline
      $(k_x^2-k_y^2,2k_xk_y)$   & 2 & + & + \\\hline
      $(k_x^3-3k_xk_y^2,
      3k_x^2k_y-k_y^3)$         & 3 & - & - \\\hline
      $(k_x^3+k_xk_y^2,
      k_x^2k_y+k_y^3)$          & 1 & - & - \\
    \hline\hline
  \end{tabular}
\end{center}
  \end{minipage}
\end{table}
From \tref{tab:symmetry}, if we hope to preserve $\trevop$, $P$ and rotation
symmetry, then up to $k^3$ the general form of the Hamiltonian is given by
\begin{eqnarray}
    &&H_0=\epsilon_{\mathbf{k}}+\mathcal{M}(\mathbf{k})\Gamma_5\nonumber\\
    &&+\mathcal{A}(\mathbf{k})
    (\Gamma_4,\Gamma_3)\left(
    \begin{array}{cc}
        \cos\theta&\sin\theta\\
        -\sin\theta&\cos\theta
    \end{array}
    \right)\left(
    \begin{array}{c}
        k_x\\k_y
    \end{array}
    \right).
    \label{eq:Ham0}
\end{eqnarray}
where $\mathcal{A}(k) = A + A_2 k^2$, $\mathcal{M}(k)= M -B k^2$ and the phase $\theta$  represents
the relative phase between $\ket{E1+}$ and $\ket{H1+}$, which can be chosen arbitrarily.
Taking $\theta=-\pi/2$, in \eref{eq:Ham0}, we recover
the BHZ Hamiltonian \cite{Bernevig06}.

We now consider additional terms which preserve rotation
symmetry and time reversal symmetry, but break the inversion symmetry. By inspecting table (\ref{tab:symmetry}),
the following three terms are possible
\begin{eqnarray}
	&H_R=\frac{\mathcal{R}(\mathbf{k})}{2}(\Gamma_{14}
    +\Gamma_{23},\Gamma_{31}+\Gamma_{24})\left(
    \begin{array}{cc}
        \cos\phi&\sin\phi\\
        -\sin\phi&\cos\phi
    \end{array}
    \right)\left(
    \begin{array}{c}
        k_x\\k_y
    \end{array}\right)\nonumber\\
    &+\frac{T_0}{2}(\Gamma_{23}-\Gamma_{14},
    \Gamma_{31}-\Gamma_{24})\left(
    \begin{array}{cc}
        \cos\psi&\sin\psi\\
        -\sin\psi&\cos\psi
    \end{array}
    \right)\left(
    \begin{array}{c}
        k_x^3-3k_xk_y^2\\3k_x^2k_y-k_y^3
    \end{array}\right)\nonumber\\
    &+S_0(\Gamma_{15},\Gamma_{25})\left(
    \begin{array}{cc}
        \cos\varphi&\sin\varphi\\
        -\sin\varphi&\cos\varphi
    \end{array}
    \right)\left(
    \begin{array}{c}
        k_x^2-k_y^2\\2k_xk_y
    \end{array}\right)
    \label{eq:HamR}
\end{eqnarray}
where $\mathcal{R}(k)=R_0+R_2k^2$.
Similar to $\theta$, the phase factors $\phi$, $\varphi$ and $\psi$
are also arbitrary and will not affect the energy spectra. If we choose
$\theta=-\frac{\pi}{2}$, $\phi=-\frac{\pi}{2}$, $\psi=\frac{\pi}{2}$
and $\varphi=\frac{\pi}{2}$, then the Hamiltonian (\ref{hgte2deg})
is recovered. This ensures that the derivation in Section 2.1 has yielded
the Hamiltonian with the correct structure.

\section{Foldy-Wouthuysen transformation of the effective HgTe Hamiltonian}
The goal of this section is to obtain an effective $2\times 2$ Hamiltonian ( where 2 stands for the spin degree of freedom) for  electron $|E1\pm\rangle$ and heavy hole $|H1\pm\rangle $ subbands including non-zero in-plane  and out-of -plane electric fields.
So far, the quantum Spin Hall effect (QSHE) was described by the Hamiltonian used in \cite{Bernevig06}, where only  the diagonal blocks of our Hamiltonian \eref{hgte2deg}
 were taken to be non-zero, i.e. for
$\efield_z,\efield_{\|} = 0$. Such a block-diagonal Hamiltonian of a HgTe QW is isomorphic to the Dirac Hamiltonian describing the relativistic motion of an
electron in two dimensions ($\op{p}_z = 0$), which couples particle and antiparticle components with the same spin direction. Here, we start from the full Hamiltonian \eref{hgte2deg}
and consider the low energy physics with the energy scale smaller than the gap $2 M$.
In this case, we can apply the perturbation formula \eref{perturbationformula}
to obtain an effective model for electron and hole subbands.
This procedure is equivalent to the Foldy-Wouthuysen (FW) transformation \cite{Foldy50},
which reduces the relativistic Dirac equation in a potential to the Pauli equation\cite{Winkler05}.
We keep terms up to linear order in the in-plane $\efield_{\|}$ and out-of-plane $\efield_z$ electric fields,
as well as the terms up to the third order in $k$. Then the effective Hamiltonians
for electron ($\hat{H}_{e}$) and hole ($\hat{H}_{h}$) subbands are given by
\begin{eqnarray}
\label{hcondband}
 &&\hat{H}_{e}  = M +  V_0(x,y) + \left(-D-B+\frac{A^2}{2M}\right) k^2  + \frac{A^2}{8M^2} e\nabla \vec{\efield}_{\|}  \nonumber \\
&& - \mathcal{R}(k) (\vecsigma \times \vec{k})_z  + \mathcal{G}(k) (e\vec{\efield}_{\|} \times \vec{k})_z \sigma_z
\\
\label{hvalband}
&&\hat{H}_{h}  =
 -M + V_0(x,y) + \left(B-D-\frac{A^2}{2M}\right) k^2  +\frac{A^2}{8 M^2} e\nabla \vec{\efield}_{\|} 			 \nonumber	 \\
&&+ \frac{1}{2}\left(\mathcal{Q}(k)\sigma_+ k_-^3 + \mathcal{Q}(k)^{\dagger} \sigma_- k_+^3\right)
- \mathcal{G}(k) (e\vec{\efield}_{\|} \times \vec{k} )_z \sigma_z  \nonumber \\
&&+ \frac{1}{2}\left(\frac{A S_0}{2 M^2} [k_-,[k_-,V_0]] \sigma_+ k_-  + h.c. \right)
\end{eqnarray}
with
\begin{eqnarray}
\mathcal{G}(k) &=& \frac{A^2}{4M^2}  \\
\mathcal{R}(k) &=& R_0  + \left(\frac{i A S_0}{M} - \frac{A^2}{4 M^2}R_0 \right) k^2
 +\frac{i A S_0}{2 M^2} e\nabla \vec{\efield}_{\|} \\
\mathcal{Q}(k) &=& i T_0 + \frac{A S_0}{M} + \frac{i A^2 R_0}{4M^2}
\end{eqnarray}

The spin-dependent term $\mathcal{G}(e\vec{\efield}_{\|} \times \vec{k})_z \sigma_z$
and the spin-independent term $\frac{A^2}{8M^2} e\nabla \vec{\efield}_{\|}$ originate directly
from the FW transformation from the Dirac type Hamiltonian in the external potential $\tilde{H}_0+V_0(x,y)$
(see equation \eref{hgte2deg}) to a Pauli type equation.
Therefore, by analogy to the relativistic electron in vacuum,
we call $\mathcal{G}(e\vec{\efield}_{\|} \times \vec{k})_z \sigma_z$
the in-plane Pauli term, while the term $\frac{A^2}{8M^2} e\nabla \vec{\efield}_{\|}$ we call the Darwin term.
Note that the Pauli term can be also visualized as resulting from a Rashba field due to the edges of a typical
mesa structure used in experiments or as coming from the atomic spin-orbit splitting but it is only active at the edges where $\vec{\efield}_{\|}$ is finite.
The Darwin term does not include a contribution from the field in z-direction due to the assumption that $\efield_z$ is constant.
The in-plane Pauli and Darwin terms appear both in the electron and hole effective Hamiltonians.
The additional terms which are proportional to $\mathcal{R}(k)$ and $\mathcal{Q}(k)$
 originate from $\tilde{H}_{\mathrm{R}}$ and are direct consequence of
the broken space inversion symmetry in the z-direction.
These  terms are usually called Rashba terms and
they give linear and cubic in k contributions for electron and heavy hole subsystems, correspondingly.
In a typical experimental setup, Rashba terms are generated by an asymmetric doping profile surrounding the quantum well and can be adjusted by a top-gate which induces a tunable electric field in z-direction. By contrast the in-plane field in Darwin and in-plane Pauli terms originate from the confining potential at the sample boundary.
\begin{figure}[tb]
 \begin{center}
 \includegraphics[width = 1\textwidth]{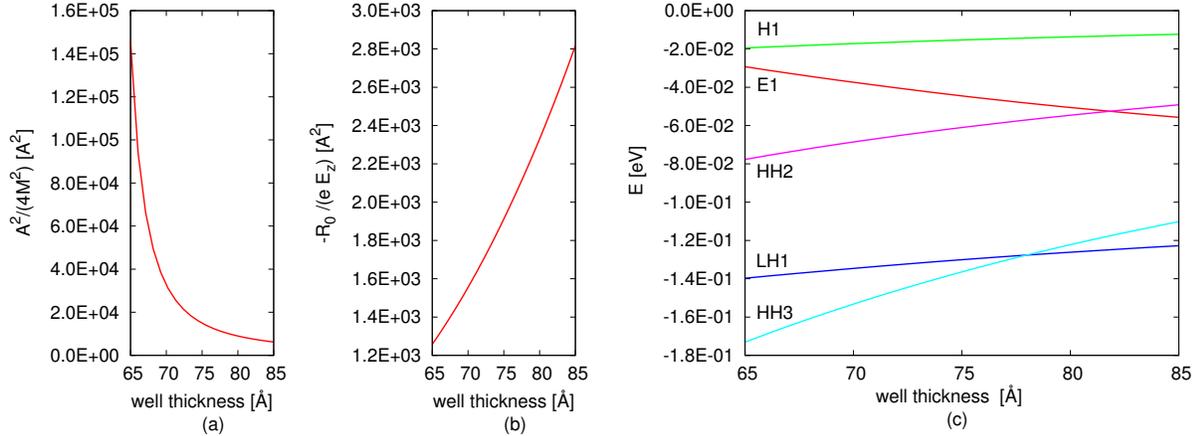}
  \caption[]{The material parameters characterizing the spin-orbit coupling induced by
(a) the in-plane field ($\frac{A^2}{4M^2}$),
(b) the out-of plane field ($R_0/eE_z$) as a function of the quantum well thickness $d$,
(c) energies of the most relevant subbands as a function of the QW width for the $\Gamma$ point.}
  \label{figurealphas}
 \end{center}
\end{figure}
Figures~\ref{figurealphas}a,b show the magnitude $\frac{A^2}{4M^2}$  of the in-plane spin-orbit interaction (SOI)
and electron Rashba coefficient $R_0/eE_z$ as a function of the  thickness $d$ of the HgTe/CdTe QW.
Note that the coupling strength $\frac{A^2}{4M^2}$ for the Pauli term decreases with $d$ while the strength of Rashba coupling $R_0/eE_z$ increases with $d$.
The origin of the different behaviours of these two SOIs can be understood from the plot of energy versus $d$ (see \fref{figurealphas}c).
The in-plane term $\frac{A^2}{4M^2}$ comes from
the coupling between the electron and the heavy hole subbands, and the energy difference between these bands increases with $d$.
The Rashba term $R_0/eE_z$ originates from the coupling between the electron
and the light hole sub-bands. Their energy difference decreases with $d$, therefore $R_0/eE_z$ increases with $d$.
Comparing the magnitudes of $R_0/eE_z$ and $\frac{A^2}{4M^2}$, one can see that close to the critical thickness $d=d_c$ determining the transition from normal to topologically non-trivial insulator, the  magnitude of the in-plane term is an order of magnitude larger than SOI term in z-direction,
while for d=80{\AA} the magnitudes of both interactions are comparable.

\section{Spin transport within an effective electron model}
As described in detail in the previous section, the effective conduction band description of a HgTe quantum well \eref{hcondband} includes two different SOI terms: the Rashba spin-orbit (SO) coupling and
the in-plane Pauli term. To understand the interplay of both SOIs, we will analyze here, the spin-Hall conductance signal numerically within the Landauer-B\"uttiker formalism \cite{Datta07}.  \\

\subsection{Description of the model}\label{subsectionModelHallBar}
The Rashba field $\efield_z$ can be applied constant in space and varied in strength easily by an external top gate. By contrast $\efield_\parallel$, generating the in plane Pauli term, usually originates from impurities or from the confinement due to sample boundaries. In our calculations we work in the quasiballistic regime, which is very well justified for HgTe/CdTe QWs with the typical mobilities 1 - 5 $\cdot$ 10$^5$ cm$^2$/Vs. Consequently the contribution of the impurities to the in-plane field is negligible and the confining potential is dominant, i.e. $\vert \efield_{x,y}\vert $ decreases with the distance from the sample edges. The confining potential requires that an electric field at the boundary always points outside the sample, i.e. its magnitude changes sign at opposite edges.

\begin{figure}[tb]
 \begin{center}
   \includegraphics[width = 0.8\linewidth]{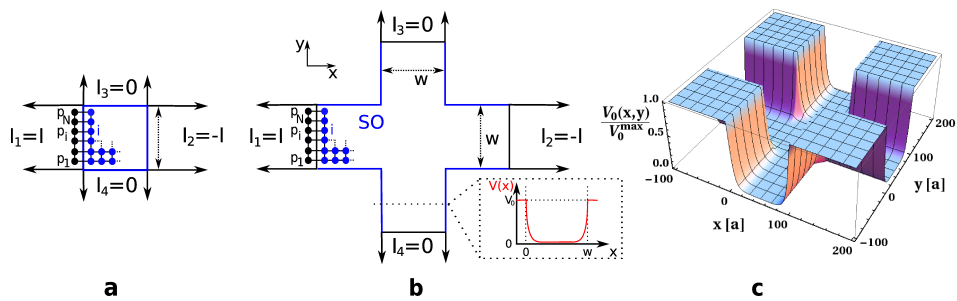}
   \caption[]{
  The two different samples we use for the numerics: (\textbf{a}) a quadratic Hall bar of width $w$ and (\textbf{b}) a cross structure. In both cases the sample (blue) contains spin-orbit interaction, while the four semi-infinite leads (black) are SOI free. The numbering of the leads and the used boundary conditions on the currents are indicated in the figure, i.e. we drive a current $I$ from lead 1 to 2 and do not allow for charge currents at the vertical leads 3 and 4. The discretization is shown for a part of each sample. Along the blue lines the confining potential $V_0(x,y)$ is applied, as indicated in the inset of (\textbf{b}), to give rise to the in-plane Pauli interaction. In case of the quadratic sample this potential corresponds to a tunneling barrier between leads and sample.
  (\textbf{c}) shows the confining potential normalized by its maximal value ($V_0^{max}$) for the cross structure.}
  \label{figureSample}
 \end{center}
\end{figure}
We use two different setups. In both setups a finite size sample with spin-orbit interactions (see figures \ref{figureSample}a,b) is attached to four semi-infinite leads of the same width $w$. For the first setup, a square sample, the in-plane field is introduced by tunneling barriers between the leads and the conductor along the blue lines in the \ref{figureSample}a. This simple setup has the advantage that the numerical results are easy to interpret, but is not a realistic  description of actual experiments. The second setup is a symmetric cross structure, which resembles the experimental Hall bars and is shown in figure \ref{figureSample}b. In this case again the blue lines indicate the sample border, where a confining potential is present (for a form of this potential see also figure \ref{figureSample}c).

We construct the confining potential in two dimensions $V_0(x,y)$ from the one dimensional profile
\begin{eqnarray}
V_{1D}(t)  = e^{-t/l} + e^{-(w-t)/l}
\end{eqnarray}
where the coordinates in both the quadratic and the cross-shaped samples are chosen such that
$0$ respectively $w$ mark the x- or y-coordinates at the edges of the central square.
$l$ is the characteristic decay length of the potential.
For the square-shaped sample, we define
\begin{eqnarray}
\label{equationEfieldExpSq}
V_0(x,y) = c_{sq} \left( V_{1D}(x) + V_{1D}(y) \right).
\end{eqnarray}
The maximal field is then $\efield_x^0 = -\partial_x V_0(x,y)|_{x=0}$. We adjust the constant $c_{sq}$ to choose some particular value $\efield_x^0$.
For the cross-shaped sample, we use the definition
\begin{eqnarray}
\label{equationEfieldExpCr}
V_0(x,y) = c_{cr} V_{1D}(\mathrm{min}(\mathrm{max}(x,0),w)) \cdot V_{1D}(\mathrm{min}(\mathrm{max}(y,0),w)).
\end{eqnarray}
Here the constant $c_{cr}$ is adjusted to choose the desired maximal field $\efield_x^0 = -\partial_x V_0(x,y)|_{x=y=0}$ and we assume that $\efield_x^0 = \efield_y^0$.
Further, in the leads the potential is always set zero.
For clarification, the confining potential corresponding to \eref{equationEfieldExpCr} is shown
in figure \ref{figureSample}c.
We find that the numerical results do not change qualitatively if the boundary field is defined differently,
as long as the characteristic decay length $l$ is unchanged.
In both setups we assume that the leads do not include SO interactions, and therefore an analytical form of the eigenmodes \cite{Sols89} and a clear definition of the spin current is available \cite{Hankiewicz04}.\\
In the calculations, we set the boundary conditions on the currents $I_1 = -I_2 = I$ and $I_3 = I_4 = 0$, where $I_p = I^\uparrow_p + I_p^\downarrow$ is the  total current at lead $p$. The spin-dependent current, $I_p^{\sigma}$, is calculated by use of the Landauer-B\"uttiker formula \cite{Buettiker85,Hankiewicz04,Datta07}
\begin{eqnarray}
  \label{equationLandauerBuettiker}
   I_p^{\sigma} = \frac{e^2}{h} \sum_{q \neq p} \sum_{\sigma^{\prime} = \uparrow, \downarrow}\left[T_{qp}^{\sigma^{\prime}\sigma} \mu_p - T_{pq}^{\sigma\sigma^{\prime}} \mu_q \right],
\end{eqnarray}
which links the spin-resolved current to the chemical potential $\mu_p  = eV_p$ via the transmission matrix elements $T_{pq}^{\sigma\sigma^\prime}$. $T_{pq}^{\sigma\sigma^\prime}$ describes the probability that an electron, entering the sample at lead $p$ with spin $\sigma$, will leave the sample through lead $q$ having spin $z$-projection $\sigma^\prime$.\\
 For a sample with non-zero SO coupling, applying an electric field between leads 1 and 2 will  generate a transverse  spin current $I_p^s = - \hbar/(2 e) (I_p^\uparrow - I_p^\downarrow)$ at leads 3 and 4, which is the so called spin-Hall effect \cite{Hankiewicz04,Nikolic05a,Hankiewicz05}.
We define the spin-Hall conductance as follows:
\begin{eqnarray}
 \label{equationSpinConductance1}
 G_p^s = \frac{I_p^s}{V_1 -V_2} = \frac{\hbar}{2 e} \frac{I_p^\uparrow- I_p^\downarrow}{V_2 - V_1}= \frac{e}{4 \pi} \sum_{\sigma^{\prime} =\uparrow, \downarrow} \left( T_{p1}^{\uparrow \sigma^{\prime}} -   T_{p1}^{\downarrow \sigma^{\prime}}\right), \quad p = 3,4
\end{eqnarray}
where $V_2 - V_1$ is the voltage difference between leads 2 and 1.
Due to the symmetries of the Hamiltonian, only a few transmission matrix elements are independent, so that the last equality in \eref{equationSpinConductance1} follows.
The time reversal symmetry of \eref{hcondband} implies $T_{pq}^{\sigma\sigma^\prime} = T_{qp}^{-\sigma^\prime -\sigma}$,
the fourfold rotational symmetry $\mathcal{C}_4$ of the setup about the $z$-axis implies e.g. $T_{23}^{\sigma \sigma'} = T_{14}^{\sigma \sigma'}$,
and the mirror symmetry with respect to the $yz$-plane implies $T_{32}^{\sigma \sigma'} = T_{31}^{-\sigma -\sigma'}$.\\
The transmission matrix elements are computed numerically in a tight binding approach
by using the Green's function method \cite{Datta07, Sols89,Hankiewicz04,Hankiewicz05}, and the Fisher-Lee relation \cite{Fisher81} connecting the Green's function with the transmission amplitudes.
We discretize the sample as indicated in Fig.~\ref{figureSample}.
By making use of the fermionic field operators $c^\dagger_{\alpha,\sigma}$ ($c_{\alpha, \sigma}$),
 which create (annihilate) a spin $\sigma$ electron at lattice site $\alpha$, the Rashba
and  in-plane Pauli interactions take the following form in second quantization:
\begin{eqnarray}
 \label{equationHamiltonianRashbaSQ}
 \fl \mathcal H_{\mathrm{Rashba}} =& \frac{R_0}{2a} \sum_{\alpha}\left[\mathrm i  c^{\dagger}_{\alpha, \uparrow} c_{\alpha+ a_y, \downarrow} +\mathrm i  c^{\dagger}_{\alpha, \downarrow} c_{\alpha+ a_y, \uparrow} -c^{\dagger}_{\alpha \uparrow} c_{\alpha + a_x \downarrow} + c^{\dagger}_{\alpha \downarrow} c_{\alpha + a_x \uparrow}   + \mathrm{h.c.} \right]
\end{eqnarray}
\begin{eqnarray}
 \label{equationHamiltonianBoundarySQ}
 \fl  \mathcal H_{\mathrm{Pauli}} =& \mathrm i\frac{A^2 e}{a 8 M^2} \sum_{\alpha,\sigma} \left[\efield_x^{\alpha+a_y/2} c^{\dagger}_{\alpha+ a_y,\sigma} c_{\alpha,\sigma} - \efield^{\alpha+a_x/2}_{y}  c^{\dagger}_{\alpha + a_x,\sigma} c_{\alpha,\sigma} + \mathrm{h.c.} \right] \kappa_{\sigma}
\end{eqnarray}
 where $R_0/(e\efield_z)$, $A$ and $M$ are the material parameters defined in section 2.1.
Here $a$ denotes the lattice constant, and $a_{x,y}$ stand for the lattice unit vectors connecting nearest neighbours.
To obtain the parameter $R_0$ we must assume a specific value of the perpendicular field $\efield_z$.
$\efield_x^{\alpha}$ and $\efield_y^{\alpha}$ designate, for a given site $\alpha$, the in-plane electric field components.
$\kappa_\sigma = \pm 1$ for spin up and down, respectively, and the symbol h.c. denotes the hermitian conjugate. \\
The Rashba spin-orbit interaction does not conserve the z-component of the spin and thus leads to spin precession \cite{Nikolic05b}.
By contrast, the in-plane Pauli term \eref{equationHamiltonianBoundarySQ} conserves the z-component of the spin, causing a shift in energy for two spin directions.
This energy shift however must not be mistaken as Zeeman effect, because \eref{equationHamiltonianBoundarySQ} does not break time reversal symmetry.\\
For numerical calculations we consider the Hamiltonian
\begin{eqnarray}
 \label{equationNumHam}
 \mathcal H = \hat T + \mathcal H_{\mathrm{SO}} + \mathcal H_{\mathrm{Dis}} + V_0(x,y) + \mathcal H_{\mathrm{Dar}}.
\end{eqnarray}
Here $\hat T =\left(-D-B+A^2/(2M)\right)\cdot k^2 = \hbar^2k^2/(2m^*)$ describes the kinetic part of the conduction band Hamiltonian \eref{hcondband}. In second quantization, $\hat T$ is described by spin-conserving nearest neighbour hopping \cite{Sols89,Datta07,Hankiewicz04}. $\mathcal H_{\mathrm{Dis}}= \mathrm{diag}(\varepsilon_i)$ specifies the disorder of the sample, where the diagonal on-site energies $\varepsilon_i$ are uniformly distributed between $[-W/2, W/2]$ \cite{Sheng05}. The disorder strength $W = \hbar e /(m^* \mu)$ is calculated from the mobility $\mu$. The confining potential is taken into account via $V_0(x,y)$. The spin-orbit coupling $\mathcal H_{\mathrm{SO}}$  is described by the Rashba term \eref{equationHamiltonianRashbaSQ}, the in-plane Pauli term \eref{equationHamiltonianBoundarySQ} or a linear superposition of both terms. Hence $\mathcal H$ mirrors the conduction band Hamiltonian \eref{hcondband}, where
we omit the k-dependence of the $\mathcal{R}$ parameter
and negligible terms which include the combined effect of the in-plane and out of plane electric fields.  The spin-independent Darwin term $\mathcal H_{\mathrm{Dar}} \propto \bi{\nabla}\cdot\efield_{\|}$ breaks the particle-hole symmetry of the tight binding Hamiltonian, just like
any space dependent in-plane potential would do. Here particle-hole symmetry means the relation $G_3^s(E_f) = -G_3^s(-E_f)$
if the energy zero point is chosen in the middle of the tight binding band.
It originates from the cosine dispersion relation of a free electron on a lattice. $H_{\mathrm{SO}}$ does not break this symmetry.
The Darwin term does not qualitatively change the spin conductance signal and will be considered after the spin-orbit terms are analysed.

We use realistic parameters for the calculations, which are shown in table \ref{tab:values}.
Here we assume a thickness of the quantum well in $z$-direction of $7\;\mathrm{nm}$, corresponding to the inverted regime. Although this has no impact in our one band approach, it guarantees a large coupling strength $R_0/(e\efield_z)$. The carrier density is set to $n = 2\cdot10^{10} \; \mathrm{cm}^{-2}$ while the effective mass $m^*=0.00712\;m_0$, where $m_0$ is the bare electron mass. Assuming a quadratic dispersion around $\bi{k} =0$, we determine the Fermi energy to be $E_f =  6.73 \;\mathrm{meV}$. Finally the assumed mobility $\mu = 25 \cdot 10^4 \; \mathrm{cm^2/(Vs)}$ leads to $W = 0.65\;\mathrm{meV}$. For such a small disorder strength averaging over 10 different disorder configurations is sufficient. We  note that the mean spin-Hall conductance deviates from that of a clean sample only by about $\lesssim 1 \%$.
The parameters are chosen carefully to the restricted range of possible energies  where this model is valid,
i.e. $E_{\mathrm r} < E_f < E_{\mathrm{gap}}$, where $E_{\mathrm{gap}} = 2 M$ is the energy gap and $E_{\mathrm r}$ is the energy splitting of the band due to Rashba interaction.

We will first focus on the square sample because the scattering barrier in this setup allows to study the competition between the in-plane and out-of-plane electric fields. The influence of in-plane electric fields is much weaker in the cross structures. Further, the first minimum in the spin-Hall conductance generated by the Rashba contribution is shifted away from the spin-precession length to smaller fields due to quantum interference effects in the vertical stubs (see e.g. \cite{Sols89}) and therefore results are less transparent to interpret.

\subsection{Numerical results for the quadratic sample}\label{subsectionNumResults}
\begin{figure}[tb]
 \begin{center}
  \includegraphics[width = 0.9 \textwidth]{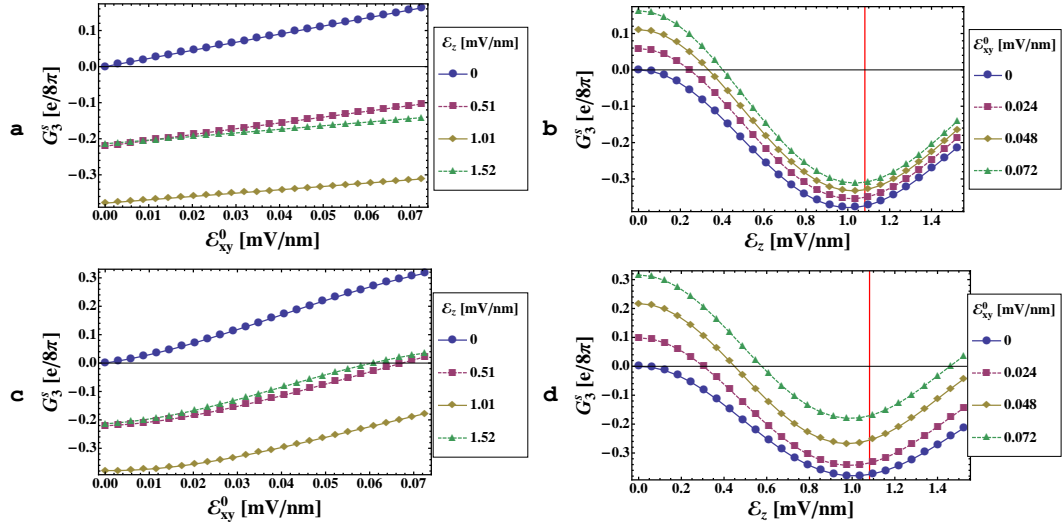}
  \caption[]{The spin-Hall conductance due to the presence of a superposition of Rashba and in-plane Pauli terms in a quadratic sample with tunneling barriers at the boundaries. Due to the presence of the second interaction the starting value of spin-Hall conductance can be nonzero. (\textbf{a}) When $\efield_z$ is nonzero, the magnitude of $G_3^s$ as function of $\efield_{x,y}^0$ is reduced. (\textbf{b}) The spin-Hall conductance as a function of $\efield_z$. For higher values $\efield_{x,y}^0$ the precession amplitude of $G_3^s$ at $L_{SO}$ (indicated by a red line) is increased and the minimum is slightly shifted to higher interaction energies. (\textbf{c}) and (\textbf{d}) show the same dependencies of $G_3^s$ as above, but here the Darwin term and the confining potential were additionally taken into account.}
   \label{figureBothSquare}
 \end{center}
\end{figure}

We choose a quadratic sample of width $w = 1000\;\mathrm{nm}$,
which is discretized by $200\times 200$ lattice points, so that the Fermi wavelength is about 36 times the lattice constant.
The characteristic length scale of the electric field is assumed to be $l = 10\;\mathrm{nm}$. The computed spin-Hall conductance is presented in figure \ref{figureBothSquare}. First we focus on pure SOI and consequently omit the Darwin term and the potential in figures \ref{figureBothSquare}a and b. Figure \ref{figureBothSquare}a shows $G_3^s(\efield_\parallel^0)$ for different top gate fields and figure \ref{figureBothSquare}b presents $G_3^s(\efield_z)$ for different in-plane fields. \\
\textbf{Rashba coupling}: The spin-Hall conductance signal induced by the Rashba coupling alone, $\efield_\parallel = 0$, is shown in figure \ref{figureBothSquare}b by blue circles. For a small interaction strength $R_0$, the spin-Hall conductance rises quadratically, saturates and finally starts
to precess. The behaviour of the spin-Hall conductance originating from the Rashba model can be understood by the spin force operator
\begin{eqnarray}
\label{equationSpinForceOperator}
 \hat{\bi F}_H = \frac{-m^*}{\hbar^2}[[\hat{\bi r}_H,\hat{H}], \hat{H}] =
 \frac{2 m^* R_0^2 }{\hbar^3}\left(\hat{\bi p}_H \times\ \bi z \right) \otimes \hat{\sigma}_H^z
\end{eqnarray}
where $\hat{\bi r}_H$, $\hat{\bi p}_H$ and $\hat{\sigma}^z_H$ are the position, momentum and spin-operators in the Heisenberg representation (see Nikoli\'c et al. \cite{Nikolic05b}) and $\bi z$ is a unit vector.
 In this simple picture the force acting on electrons due to SO coupling is quadratic in $R_0$,
explaining the behaviour of $G_3^s$ as a function of the out of plane electric field for low $\efield_z$.
The force described in \eref{equationSpinForceOperator} deflects the spin-$\uparrow$ and the spin-$\downarrow$ electrons in  opposite transverse directions leading to the spin-Hall effect. However, the Hamiltonian does not conserve the z-component of the spin,
leading to a rotation of the spin direction and as a consequence to oscillations in $G_3^s$ as a function of $R_0$. The first maximum of the spin-Hall conductance is reached, when the spin has travelled a distance equivalent to the spin precession length \cite{Nikolic05a}
 $L_{SO} = \pi \hat T/(k^2 R_0)$, over which the expectation value $\langle \sigma_z \rangle$ rotates by $\pi$.
 The electric field $\efield_z^{SO}= 1.08 \;\mathrm{mV/nm}$
 corresponding to the precession length is indicated as a red line in figure \ref{figureBothSquare}b and is in good agreement with the maximum of the absolute value of $G_3^s$. \\
\textbf{In-plane Pauli interaction}: The spin-Hall conductance shows a linear behaviour as a function of the in-plane electric field (see blue circles in
 figure \ref{figureBothSquare}a, where only the $\efield_{x,y}$ components are nonzero).
This linear dependence on field strength can be explained within the semiclassical approach, where we have adopted the wave packet dynamics by Sundaram and Niu \cite{Sundaram99} to obtain equations of motion for the in-plane Pauli term:
\numparts
\begin{eqnarray}
 \label{equationSemiclassical1}
  \dot{\bi{r}}_c &= \frac{1}{\hbar} \frac{\partial \epsilon}{\partial \bi{k}_c} - \dot{\bi{k}}_c \times \bi{\Omega}_\sigma \\
 \label{equationSemiclassical2}
  \hbar \dot{\bi{k}}_c &= - e \efield_{\|},
\end{eqnarray}
\endnumparts
with the magnetic field set to zero.
Here the index $c$ denotes the center coordinate of the wave packet in position and $\bi{k}$ space. The Berry curvature is defined as
\begin{equation}
 \label{equationBerryCurvature}
 \left(\bi{\Omega^{\pm}}_\sigma\right)_\alpha := -\varepsilon_{\alpha\beta\gamma} \mathrm{Im} \left\langle \frac{\partial u^{\pm}_\sigma(\bi{k})}{\partial k_\beta}  \left\vert  \frac{\partial u^{\pm}_\sigma(\bi{k})}{\partial k_\gamma} \right.\right\rangle,
\end{equation}
where symbol $\pm$ corresponds to two eigenvalues $E_{\pm} =-Dk^2\pm \sqrt{A^2k^2+\mathcal{M}^2(k)}$ of the upper spin block of the Hamiltonian  $\tilde{H}_0$ in \eref{H0} with $u^{\pm}_\sigma (\bi k)$ being corresponding eigenstates for spin $\sigma$.
\Eref{equationSemiclassical2} simply describes the change of the lattice momentum due to the electric field.
\Eref{equationSemiclassical1} describes the time evolution of the position operator  due to the band dispersion and the anomalous velocity term
$\dot{\bi{k}}_c \times \bi{\Omega}_\sigma$ with the spin-dependent non-zero z-components  $(\bi \Omega^{\pm}_\uparrow)_z = - (\bi \Omega^{\pm}_\downarrow)_z\stackrel{\bi k = 0}{=}\pm A^2/2M^2$.
 The spin-dependent anomalous velocity term shifts the position of the wave packet with different spins in two opposite
 transverse directions leading to the spin-Hall effect.
 Inserting \eref{equationSemiclassical2} into \eref{equationSemiclassical1} yields the dependence of the anomalous velocity term linear in $\efield_{\|}$.
 Note, that the energy range ($\approx 0.011 t$) plotted in figures \ref{figureBothSquare}a and b is the same for both interaction terms. Since the coupling parameters can be quite different (in our calculations $21\cdot R_0/(e\efield_z) \approx  A^2/(4M^2)$),
 the magnitudes of in-plane and out-of-plane electric fields are adjusted so the interaction energies are the same.\\
\textbf{Interplay of both interactions}: A linear superposition of \eref{equationHamiltonianRashbaSQ}, \eref{equationHamiltonianBoundarySQ} and $\hat T$ leads to the spin signal which is shown in figures \ref{figureBothSquare}a and b, when all three field components are nonzero.
The finite value of the spin-Hall conductance in figure \ref{figureBothSquare}a for $\efield_{x,y}^0 = 0$ is due to the Rashba coupling. It can be observed, that the linear behaviour of the in-plane Pauli term with the electric field  is not changed, when Rashba spin-orbit coupling is present.
However, the slope of the spin-Hall conductance curves decreases with $\efield_z \neq 0$, which means that the in-plane Pauli contribution to the spin signal is suppressed by the Rashba interaction. The z-component of the spin is not conserved for finite $\efield_z$, as can be seen from equation \eref{equationHamiltonianRashbaSQ}. The resulting spin precession implies that generation of a spin current by the anomalous velocity becomes less effective.
The smallest slope and therefore smallest in-plane Pauli contribution in figure \ref{figureBothSquare}a is found for an electric field  $\efield_z$ corresponding to the precession length, where an expectation value $\langle \sigma_z \rangle$ rotates by $\pi$.

Figure \ref{figureBothSquare}b shows $G_3^s$ as a function of $\efield_z$ for different in-plane electric fields.
One can see the typical precession pattern of $G_3^s(\efield_z)$ also for $\efield_\parallel \neq 0$.
Moreover the precession amplitude of the spin-Hall conductance of the Rashba type is enhanced in the presence of the in-plane Pauli term.
The origin of this increase can be traced
back to the $\bi{k}$-dependent energy splitting of the spin subbands due to the in-plane Pauli interaction. In order to lower its energy the electron now prefers to stay in either spin up or down states. The precession of the spin is thus slightly suppressed, as can be seen in a small shift of the minima to higher electric fields. As discussed above, the spin force operator \eref{equationSpinForceOperator} can act more efficiently on electrons with a preferred spin $z$-projection, which leads to the relatively higher magnitude of the spin-Hall conductance caused by Rashba coupling.\\
\Fref{figureBothSquare}c and \Fref{figureBothSquare}d  also show $G_3^s(\efield_\parallel)$ and $G_3^s(\efield_z)$ respectively but now with included potential $V_0(x,y)$ and  Darwin $A^2/(8M^2) \bi{\nabla} \cdot \efield_{\parallel}$ terms. In the tight binding approach, both terms renormalize the diagonal on-site energy. They cannot generally be considered to be small, as they scale, like the in-plane Pauli term, with the magnitude and the shape of the confining electric field. The relative magnitudes of the in-plane Pauli, Darwin and potential terms depend on the choice of the functional dependence $V_0(x,y)$. The most important scale is the characteristic length scale $l$, over which the corresponding field drops to $\efield^0/e$. We have performed numerical calculations with different values $\efield^0_{x,y}$ and $l$ and found, that the main features of the results discussed in this paper stay the same. For the configuration we choose the main renormalization in respect to figures \Fref{figureBothSquare}a and \Fref{figureBothSquare}b is due to the Darwin term. \\
The divergence of $\vec{\efield}_{\|}$ appears as an additional term in the semiclassical equation \eref{equationSemiclassical2} and therefore the spin-independent Darwin term can contribute to the anomalous velocity and renormalize the spin-Hall conductance term. This can be seen in figure \ref{figureBothSquare}c as a non linear behaviour of $G_3^s (\efield_\parallel)$.
 However, in the range of in-plane electric fields shown in figures 3c and 3d, the qualitative behaviour of the spin-Hall conductance is the same as in the absence of the Darwin term (see \fref{figureBothSquare}a,b).
  Increasing the in-plane electric field to the same magnitude as the electric field perpendicular to the 2DEG bears two difficulties. First of all the interaction energy of the in-plane Pauli term exceeds the Fermi energy which marks the limit of validity of our effective electron model. Secondly increasing $\efield_\parallel$ comes along with a raising tunneling barrier in the quadratic sample.  We can omit these difficulties by choosing the sample in a shape of a cross (see next subsection).

\subsection{Numerical results for the cross sample}\label{subsectionNumResults2}
\begin{figure}[tb]
 \begin{center}
  \includegraphics[width = 0.9\textwidth]{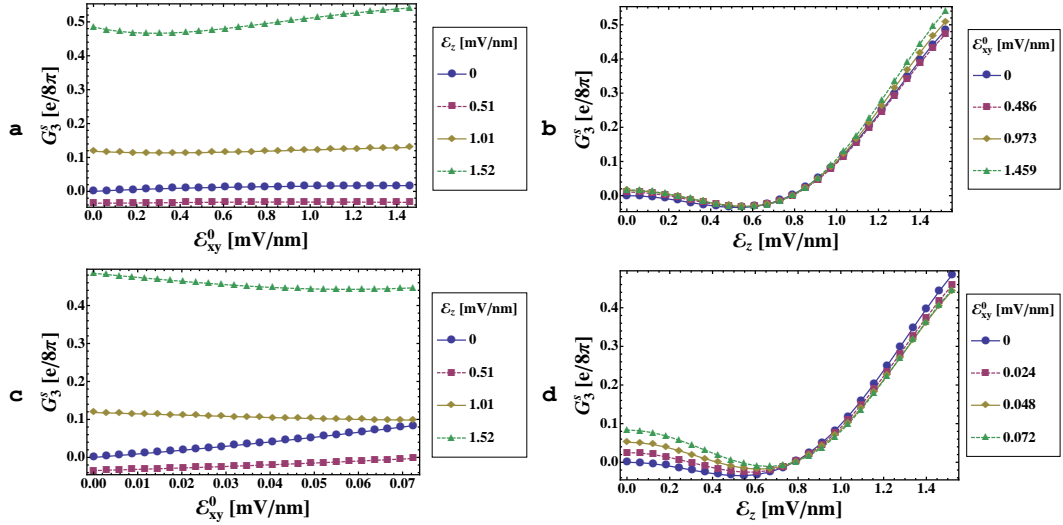}
  \caption[]{The spin-Hall conductance for a cross structure, with the  Darwin and the potential terms included. In (\textbf{a}) and (\textbf{b}) we have used the same range of the electric field as for the quadratic sample, i.e. $l = 10\; \mathrm{nm}$. Although the in-plane field was enhanced to very high values, the influence of in-plane electric field on $G_3^s(\efield_z)$ is very weak. In (\textbf{c}) and (\textbf{d}) we have increased $l$ to $40\;\mathrm{nm}$ to get a higher in-plane Pauli signal. In this case $G_3^s$ behaves similarly as for the quadratic sample.}
  \label{figureCrossStructure}
 \end{center}
\end{figure}

The cross sample is constructed of 5 square parts: 4 stubs and the central square (see figure \ref{figureSample}b). Each part has the width $w = 500\;\mathrm{nm}$ and is discretized by $100\times 100$ lattice sites. The corresponding spin-Hall conductance  originating from in-plane Pauli and Rashba terms in the presence of the Darwin term and the confining potential is shown in figures \ref{figureCrossStructure}a - d. \\
Figures \ref{figureCrossStructure}a,b show $G^s_3(\efield_\parallel)$ and $G^s_3(\efield_z)$ for different  values of fields in z and in-plane directions correspondingly and for the characteristic range of the electric field $l=10$nm.
These figures should be compared with figures \ref{figureBothSquare}c and \ref{figureBothSquare}d correspondingly.
The overall behaviour  of spin-Hall conductance in figures \ref{figureCrossStructure}a and b is similar to the quadratic sample. However, although we used values of in-plane electric field  around twenty times larger  than in figure \ref{figureBothSquare}d the influence of  $\efield_\parallel$ on $G^s_3(\efield_z)$ is much weaker  than for the square structure (compare \ref{figureCrossStructure}b with \ref{figureBothSquare}d ).
 Probability of scattering from  vertical and horizontal walls in the cross structure is much smaller
 than in a case of square structure where electron directly hits the wall.
Therefore we find that the influence of in-plane electric field will be much weaker in the experimentally relevant cross samples.

In order to obtain a contribution to $G_3^s$ due to the confining potential comparable to that for the quadratic sample, we have increased $l$ from $10\;$nm to $40\;$nm, in figures \ref{figureCrossStructure}c,d. This leads to a larger range of lattice sites, which can contribute efficiently to the spin dependent hopping.
In the case of the larger $l$ the results resemble those discussed in the last section for square samples. In figure \ref{figureCrossStructure}d the amplitude of the Rashba contribution to the spin-Hall effect counted to the first minimum is enhanced by the confinement potential. By contrast the signal due to the in-plane Pauli term decreases until $\efield_z$ corresponds to the first minimum as can be seen in figure \ref{figureCrossStructure}c. The magnitude of the slope increases slightly for higher values of the top gate field, but with an inverted sign with respect to the quadratic sample (see figure \ref{figureBothSquare}c).  \\

At the end of this section let us emphasize, that for the experimentally relevant case both interactions are present,
but only $\efield_z$ can be easily varied, e.g. by a top gate.  Therefore, in the experimentally relevant case, the presence of both the in-plane Pauli and the Darwin interactions could lead to an increase of the amplitude of spin signal.

\section{Conclusions}
We have derived an extension to the BHZ Hamiltonian for the typical 2D topological insulator (HgTe QWs) in the presence of the
inversion breaking potential in z-direction and in-plane confining potential. For the derivation, we used  two independent
methods: $\vec{k}$$\cdot$$\vec{p}$ perturbation theory and symmetry arguments based on Clifford algebra. We found that to the
third order in the perturbation theory, only the inversion breaking potential in z-direction generates new off-diagonal in spin space terms.
These terms lead to the Rashba spin-orbit  interaction when the  Foldy-Wouthuysen  transformation to the effective electron model is performed. On the other hand the diagonal-in-spin space part of the Hamiltonian in the presence of the confining in-plane potential generates  an additional term to the one band model which is also linear in momentum and spin, but conserves the z-component of the spin. By analogy with the equation for a relativistic electron in vacuum we call  this term in-plane Pauli term. The presence of both terms in the conduction band Hamiltonian leads to an interesting behaviour of the spin-Hall conductance. In particular, the in-plane Pauli contribution to the spin-Hall conductance is suppressed in the presence of the spin precession inducing terms.
 By contrast, the spin-Hall conductance from the Rashba term  preserves the oscillation pattern in the presence of the in-plane Pauli term and its magnitude can be enhanced  due to partial pinning of the z-component of the spin.
This latter situation is experimentally relevant since the inversion breaking potential in z-direction can be easily tuned by a top gate in experiments.
Therefore we expect that in experiments on asymmetrically doped HgTe/CdTe QWs \cite{Bruene08} in the metallic regime (the Fermi level in the conduction or valence band), the behaviour of spin transport and especially the spin-Hall conductance will be dominated by the Rashba spin-orbit interaction.
Note, that in our derivation we omit the BIA terms since they are already studied in Ref. \cite{Koenig07}.

 Let us also emphasize that our effective four band Hamiltonian in the presence of inversion breaking potential is not limited only to the HgTe/CdTe QWs and can be  easily generalized to other topological insulators like type II InAs/GaSb/AlSb quantum wells\cite{Liu08} or Bi$_2$Se$_3$ thin film\cite{Liu10,Linder10,Lu2009} with correctly adjusted strengths of the Rashba spin-orbit interactions.

\section{Acknowledgements}
We gratefully acknowledge the support by the German
DFG grant no. HA5893/1-1. C.-X. Liu acknowledges
financial support
by the Alexander von Humboldt Foundation of Germany.
LWM acknowledges the joint DFG-JST Forschergruppe on Topological Electronics and
German-Israeli Foundation grant I-881-138.7/2005.
SCZ is supported by the US
NSF under grant number DMR-0904264.
We thank Leibniz Rechenzentrum Munich for
providing computing resources.

\bibliographystyle{unsrt}

\begin{thebibliography}{10}

\bibitem{Bernevig06}
B.~A. Bernevig, T.~L. Hughes, and S.-C. Zhang.
\newblock {\em Science}, 314:1757, 2006.

\bibitem{Koenig07}
M.~K\"onig, S.~Wiedmann, C.~Br\"une, A.~Roth, H.~Buhmann, L.~W. Molenkamp,
  X.-L. Qi, and S.-C. Zhang.
\newblock {\em Science}, 318:766, 2007.

\bibitem{Koenig08}
M.~K\"onig, H.~Buhmann, L.~W. Molenkamp, T.~Hughes, C.-X. Liu, X.-L. Qi, and
  S.-C. Zhang.
\newblock {\em J. of the Phys. Soc. of Japan}, 77(3):031007, 2008.

\bibitem{Roth09}
A.~Roth, C.~Br\"une, H.~Buhmann, L.~W. Molenkamp, J.~Maciejko, X.-L. Qi, and
  S.-C. Zhang.
\newblock {\em Science}, 325:294, 2009.

\bibitem{Fu07}
L.~Fu and C.~L. Kane.
\newblock {\em Phys. Rev. B}, 76:045302, 2007.

\bibitem{Zhang09}
H.~Zhang, C.-X. Liu, X.-L. Qi, X.~Dai, Z.~Fang, and S.~C. Zhang.
\newblock {\em Nature Physics}, 5:438, 2009.

\bibitem{Novik05}
E.~G. Novik, A.~Pfeuffer-Jeschke, T.~Jungwirth, V.~Latussek, C.~R. Becker,
  G.~Landwehr, H.~Buhmann, and L.~W. Molenkamp.
\newblock {\em Phys. Rev. B}, 72(3):035321, Jul 2005.

\bibitem{Bruene08}
C.~Br\"{u}ne, A.~Roth, E.~G. Novik, M.~Koenig, H.~Buhmann, E.~M. Hankiewicz,
  W.~Hanke, J.~Sinova, and L.~W. Molenkamp.
\newblock {\em e-print arXiv: 0812.3768}, 2008.

\bibitem{Koenig06}
M.~K\"onig, A.~Tschetschetkin, E.~M. Hankiewicz, J.~Sinova, V.~Hock, V.~Daumer,
  M.~Sch\"afer, C.~R. Becker, H.~Buhmann, and L.~W. Molenkamp.
\newblock {\em Phys. Rev. Lett.}, 96:076804, 2006.

\bibitem{Liu08}
C.-X. Liu, T.~L. Hughes, X.-L. Qi, K.~Wang, and S.-C. Zhang.
\newblock {\em Phys. Rev. Lett.}, 100:236601, 2008.

\bibitem{Liu10}
C.-X. Liu, H.J. Zhang, B.~Yan, X.-L. Qi, T.~Frauenheim, X.~Dai, Z.~Fang, and
  S.-C. Zhang.
\newblock {\em Phys. Rev. B}, 81:041307, Jan 2010.

\bibitem{Linder10}
J.~Linder, T.~Yokoyama, and A.~Sudb\o{}.
\newblock {\em Phys. Rev. B}, 80:205401, Nov 2009.

\bibitem{Lu2009}
H.-Z. Lu, W.-Y. Shan, W.~Yao, Q.~Niu, and S.-Q. Shen.
\newblock arxiv: cond-mat/0908.3120, 2009.

\bibitem{Winkler05}
R.~Winkler.
\newblock {\em Spin-Orbit Coupling Effects in Two-Dimensional Electron and Hole
  Systems}.
\newblock Springer, 2005.

\bibitem{Burt88}
M.~G. Burt.
\newblock {\em Semicond. Sci. Technol.}, 3:739, 1988.

\bibitem{Luttinger56}
J.~M. Luttinger.
\newblock {\em Phys. Rev.}, 102(4):1030--1041, May 1956.

\bibitem{Ekenberg85}
U.~Ekenberg and M.~Altarelli.
\newblock {\em Phys. Rev. B}, 32(6):3712--3722, Sep 1985.

\bibitem{Jeschke00}
A.~Pfeuffer-Jeschke.
\newblock PhD thesis, Universit\"at W\"urzburg, 2000.

\bibitem{Foldy50}
L.~L. Foldy and S.~A. Wouthuysen.
\newblock {\em Phys. Rev.}, 78:29, 1950.

\bibitem{Datta07}
S.~Datta.
\newblock {\em Electronic Transport in Mesoscopic Systems}.
\newblock Cambridge University Press, 2007.

\bibitem{Sols89}
F.~Sols, M.~Macucci, U.~Ravaioli, and K.~Hess.
\newblock {\em J. Appl. Phys.}, 66(8):3892, 1989.

\bibitem{Hankiewicz04}
E.~M. Hankiewicz, L.~W. Molenkamp, T.~Jungwirth, and J.~Sinova.
\newblock {\em Phys. Rev. B}, 70:241301(R), 2004.

\bibitem{Buettiker85}
M.~B\"uttiker, Y.~Imry, R.~Landauer, and S.~Pinhas.
\newblock {\em Phys. Rev. B}, 31(10):6207--6215, May 1985.

\bibitem{Nikolic05a}
B.~K. Nikoli\ifmmode~\acute{c}\else \'{c}\fi{}, L.~P. Z\^arbo, and S.~Welack.
\newblock {\em Phys. Rev. B}, 72(7):075335, Aug 2005.

\bibitem{Hankiewicz05}
E.~M. Hankiewicz, J.~Li, T.~Jungwirth, Q.~Niu, S.-Q. Shen, and J.~Sinova.
\newblock {\em Phys. Rev. B}, 72:155305, 2005.

\bibitem{Fisher81}
D.~S. Fisher and P.~A. Lee.
\newblock {\em Phys. Rev. B}, 23(12):6851--6854, Jun 1981.

\bibitem{Nikolic05b}
B.~K. Nikoli\'{c}, L.~P. Z\^arbo, and S.~Souma.
\newblock {\em Phys. Rev. B}, 72(7):075361, Aug 2005.

\bibitem{Sheng05}
L.~Sheng, D.~N. Sheng, and C.~S. Ting.
\newblock {\em Phys. Rev. Lett.}, 94(1):016602, Jan 2005.

\bibitem{Sundaram99}
G.~Sundaram and Q.~Niu.
\newblock {\em Phys. Rev. B}, 59(23):14915--14925, Jun 1999.

\end{thebibliography}

\end{document}